\documentclass[aps,prx,twocolumn,amsmath,amssymb,superscriptaddress,groupedaddress]{revtex4}  
\usepackage{graphicx}  
\usepackage{bm}        
\usepackage{amsmath}
\usepackage{amssymb}   
\usepackage{color}
\usepackage{ulem}
\usepackage{multirow}
\usepackage{xcolor}
\usepackage{mathtools}

\begin{document}
\title{Scale-free switching of polarization in the layered ferroelectric material CuInP$_2$S$_6$}

\author{N.~Sivadas} 
\email{n.sivadas@samsung.com}
\affiliation{Center for Nanophase Materials Sciences, Oak Ridge National Laboratory, Oak Ridge, Tennessee 37831, USA}
\affiliation{Advanced Materials Lab, Samsung Advanced Institute of Technology-America, Samsung Semiconductor, Inc., Cambridge, Massachusetts 02138, USA}

\author{Bobby G. Sumpter}
\affiliation{Center for Nanophase Materials Sciences, Oak Ridge National Laboratory, Oak Ridge, Tennessee 37831, USA}

\author{P. Ganesh}
\email{ganeshp@ornl.gov}
\affiliation{Center for Nanophase Materials Sciences, Oak Ridge National Laboratory, Oak Ridge, Tennessee 37831, USA}

\date{\today}

\begin{abstract}
Using first-principles calculations we model the out-of-plane switching of local dipoles in CuInP$_2$S$_6$ (CIPS) that are largely induced by Cu off-centering.  Previously, a coherent switching of polarization via a quadruple-well potential was proposed for these materials. In the super-cells we considered, we find multiple structures with similar energies but with different local polar order. Our results suggest that the individual dipoles are weakly coupled in-plane and under an electric field at very low temperatures these dipoles in CIPS should undergo incoherent disordered switching. The barrier for switching is determined by the single Cu-ion switching barrier. This in turn suggests a scale-free polarization with a switching barrier of $\sim$ 203.6-258.0 meV, a factor of five smaller than that of HfO$_2$ (1380 meV) a prototypical scale-free ferroelectric. The mechanism of polarization switching in CIPS is mediated by the switching of each weakly interacting dipole rather than the macroscopic polarization itself as previously hypothesized. These findings reconcile prior observations of a quadruple well with sloping hysteresis loops, large ionic conductivity even at 250~K well below the Curie temperature (315~K), and a significant wake-up effects where the macroscopic polarization is slow to order and set-in under an applied electric field. We also find that computed piezoelectric response  and the polarization show a linear dependence on the local dipolar order. This is consistent with having scale-free polarization and other polarization-dependent properties and opens doors for engineering tunable metastability by-design in CIPS (and related family of materials) for neuromorphic applications.



\end{abstract}
\maketitle

\section{Introduction}

Materials that exhibit scale-free ferroelectricity with atomically thin domain walls are rare. In such materials the barrier to overcome the flip of one dipole moment is the same as the barrier for a uniform switching of dipoles. Typically, this leads to multiple polar states that are stable, a property that has been sought after in memristors with applications in neuromorphic computing. HfO$_2$ is one such example that has attracted a lot of attention recently. Unlike conventional ferroelectrics (FEs) like PbTiO$_3$, HfO$_2$ exhibits sharp domain walls  with localized dipoles~\cite{Lee20p1343}. In HfO$_2$, the polarization forms in two-dimensional slices separated by nonpolar spacers. The reported switching barrier for polarization in HfO$_2$ is 1380~meV ~\cite{Lee20p1343}. Identifying materials where the switching barrier is large enough to have a scale-free polarization, yet small enough to be overcome by an electric field is a crucial next step in advancing low-power  fast microelectronic devices.

In this pursuit, two-dimensional (2D) materials with switchable polarization are an exciting alternative. CuInP$_2$S$_6$ (CIPS) is a promising van der Waals (vdW) material with an out-of-plane switchable polarization at room temperatures down to the ultra-thin-film limit~\cite{Liu16p12357}. It undergoes an order-disorder transition at $T_c$ $\sim$ 315~K, from a high-temperature paraelectric (PE) $\sl {C2/c}$ phase to a low-temperature ferroelectric phase with space-group $\sl {Cc}$~\cite{Vysochanskii98p9119}. It also hosts a negative piezoelectric response~\cite{Liu17p207601,Neumayer19p024401,You19p3780,Qi21p217601} along with a negative capacitance state ~\cite{Neumayer20p2001726}.  CIPS is regarded as a molecular ferroelectric comprised of individual Cu/In intercalated P$_2$S$_6$ polar molecular units. Given these facts, it is reasonable to ask whether CIPS behaves as a conventional ferroelectric where the switching is mediated by the coherent motion of atomic displacements or should it behave as a collection of individual dipolar molecules that switches almost independently on application of an external electric field. We will refer to the latter case as incoherent switching.



In this article, we answer this using first-principles density functional theory calculations. We investigate the degree of anharmonicity stabilizing the local dipoles, nature of the coupling between local dipoles and the barrier for switching local dipoles. For scale-free bulk polarization we expect he total polarization is largely an arithmetic sum of individual dipoles. And, for incoherent switching we expect the barrier to flip a single dipole is similar to that of switching the collective polar order. In CIPS, we find both these to be true. We find a large degree of anhamonicity in CIPS between the polar-mode and a centrosymmetric Raman active mode, similar to our findings in a related CuInP$_2$Se$_6$~\cite{Sivadas22p013094}. In addition, we find multiple metastable structures with total energies that differ by less than 20 meV/f.u. using a $2 \times 2$ unit-cell, indicating that local dipoles are weakly coupled. The barrier for flipping these local dipoles via the octahedral center are similar (203.6-258.0 meV) for these different configurations. Our model suggests that the dipole moments in CIPS behaves as non-interacting localized dipoles, which we subsequently demonstrate using {\sl ab initio} molecular dynamics (AIMD). Individual motion of Cu-atoms at low temperatures and fields explains experimentally observed unexplained low-temperature ionic conductivity down to T=250~K, well below T$_c$, particularly when Cu-deficiency is present in the system~\cite{Mat-Horiz}. The ability to form innumerable disordered metastable dipole configurations should also lead to wake-up effects, as seen in dipolar glassy systems.  We also report the corresponding piezoelectric response values ($d_{zz}$) for the different structures considered as it is the most popular way for characterizing the polar phases~\cite{Brehm20p43}, and from a thermodynamic phase-diagram demonstrate how incoherent switching will lead to a sloping hysteresis loop as observed in prior experiments. This could spur new experiments to verify our predicted scale-free switching of CIPS, and incorporate CIPS in novel neuromorphic device geometries. 


\section{Theory}

\begin{figure}
\includegraphics[width=1.0\columnwidth]{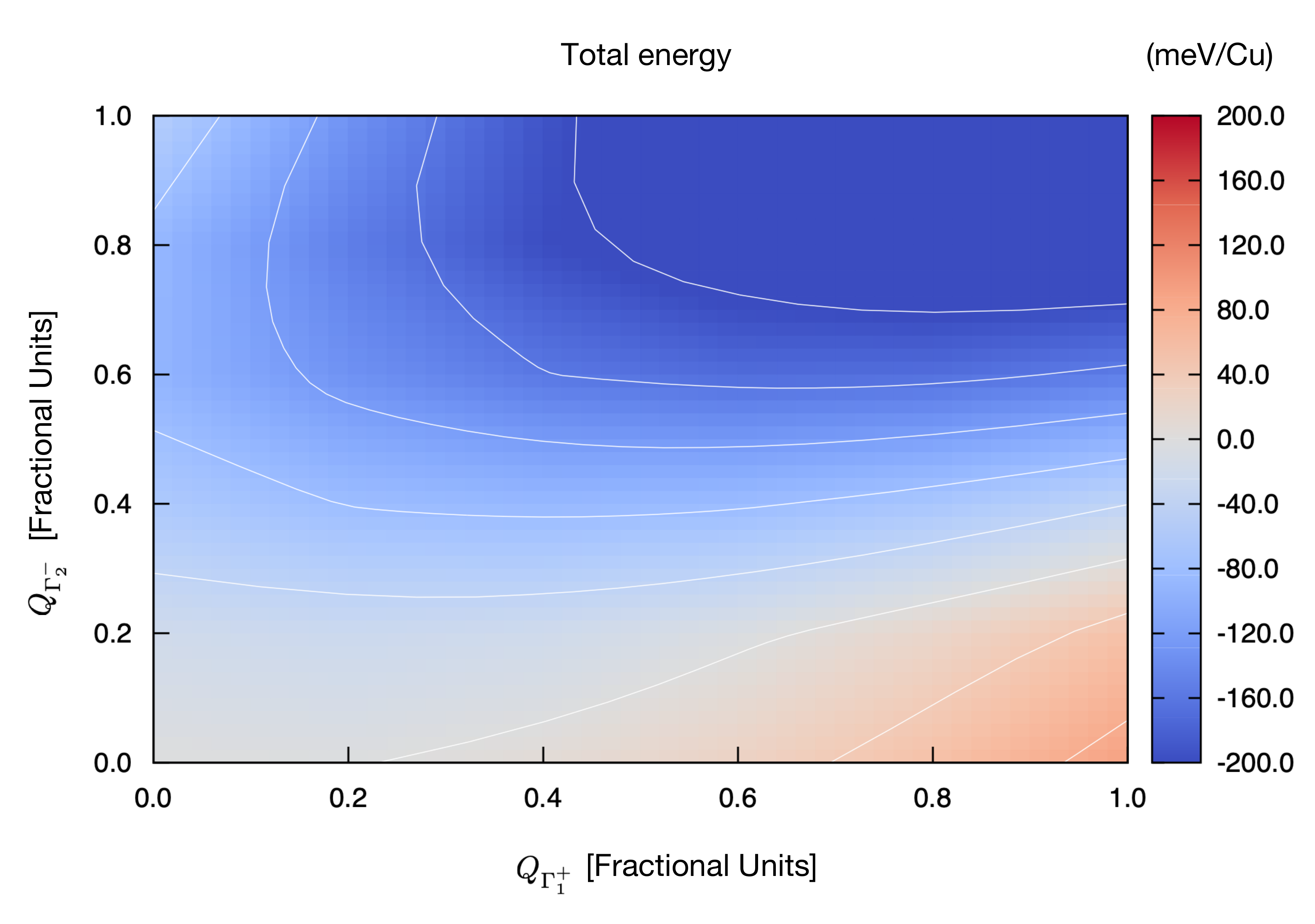}
\caption{\label{Grid} The total energy (meV/f.u.) as a function of the
fractional amplitude of the polar mode ($Q_{\Gamma_2-}$) and the fully symmetric Raman mode ($Q_{\Gamma_1+}$).}
\end{figure}

\begin{figure}
\includegraphics[width=1.0\columnwidth]{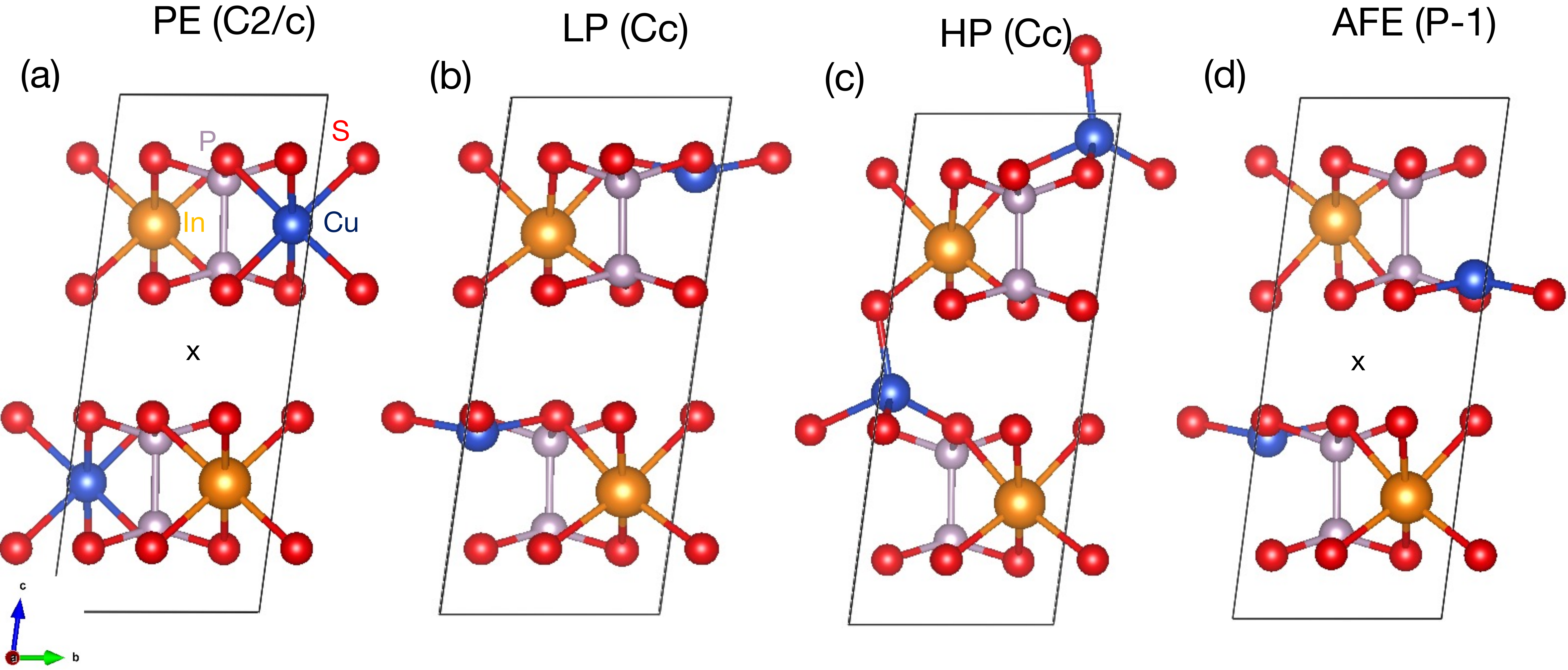}
\caption{\label{Fig1} The side view of the crystal structure of (a) the paraelectric (PE), (b) the low-polarization (LP) phase, (c) the high-polarization (HP) phase, and (d) the interlayer antiferroelectric (AFE) phase. The space group in each case is labeled. The interlayer inversion center is marked by an `$\times$'. }
\end{figure}

In our previous work, we discussed in detail that the polarization in this family of materials is stabilized by a sizable on-site anharmonic coupling between the polar mode and the non-polar Raman active modes~\cite{Sivadas22p013094}. Fig.~\ref{Grid} shows the total energy in meV per f.u. for the LP phase shown in Fig.~\ref{Fig1} (b) as a function of the fractional amplitude of the Raman active modes and the polar mode relative to the PE phase (Fig.~\ref{Fig1} (a)). The energy surface relative to the PE phase is anharmonic with a large anharmonic coupling between the polar mode and the fully-symmetric Raman active mode, similar to the Selenides~\cite{Sivadas22p013094}. However, unlike in the case of the Selenides, the polar mode alone creates a double well with additional energy gain coming from the anharmonic coupling between the polar mode and the Raman active mode. The total energy difference between the LP phase and the PE phase is 252.2 meV/f.u..

We map the polar energy surface to a classical anharmonic 1D oscillator model for dipoles~\cite{Aubry75p3217}. The potential $V$ with respect to the ordered PE phase as the reference structure can be written as, 
\begin{equation}\label{Eq1}
V = \sum_{n} \underbrace{(A x_n^2 + B x_n^4)}_{V_{on-site}} + \sum_{n,m} C (x_n -x_m)^2,
\end{equation}
where $x_n$ corresponds to the polar order parameter. This can be approximated in CIPS as the displacement of the Cu atoms. So, $x_n$ is $\pm$1 depending on the local polar site, and zero at the octahedral center. This term will be zero in the ordered PE phase. This model assumes that the on-site potential ($V_{on-site}$) represented by the first two terms with coefficients $A$ and $B$ in Equ.~\ref{Eq1} is the same for all ordering of dipoles. The last term ($C$) captures the coupling between the different local polar sites. For conventional FEs, the condensation of local dipoles will nucleate more dipoles, leading to a uniform switching. Here, we expect $C$ to be comparable to $V_{on-site}$. However, to get scale-free polarization we need $C$ to be much smaller than $V_{on-site}$ so that each dipole switch independently and without any additional energy penalty. This will lead to a combinatorially large number of metastable states, which is only limited by the number of formula-units considered to make the FE device, thereby achieving ideal memristive behavior in the 2D limit~\cite{Marinella21p36}.


\section{First-principles methodology}

We calculated the total energies using first-principles calculations as implemented in Vienna \textit{ab~initio} simulation package (VASP)~\cite{Kresse96p11169}, with the PBE functional. Structural relaxation was done with a force convergence tolerance of 0.1 meV/\AA using a conjugate-gradient algorithm. The convergence criterion for the electronic self-consistent calculations was set to 10$^{-8}$~eV. A regular 8 $\times$ 8 $\times$ 4 $\Gamma$-centered $k$-point grid was used to sample the Brillouin zone for the LP phase with a plane-wave cutoff energy of 600~eV. The computed lattice parameters and the occupied Wyckoff positions of the fully relaxed LP phase agree well with the reported experimental parameters (see supplementary)~\cite{Maisonneuve97p10860}. The total polarization was computed using the Berry-phase approach~\cite{King-Smith93p1651} where the center of mass of the P atoms was chosen as the origin. We used the ISOTROPY software suite to aid with the group-theoretic analysis~\cite{isotrophy}. The framework used to compute the piezoelectric response is similar to that which was implemented by Kim et al.~\cite{Kim19p104115}. See supplementary for a more detailed discussion.

 The computed polarization of 3.11~$\mu C/cm^2$ with DFT-D2 correction~\cite{Grimme06p1787} for the LP phase compares well with the values (3.04~$\mu C/cm^2$) reported by Qi et al.~\cite{Qi21p217601} who also used the DFT-D2 correction. For DFT-D2, we also find $e_{zz}$ to be -10.8~$\mu C/cm^2$, similar to what was reported by Qi et al.~\cite{Qi21p217601}. However, as DFT-D2 does not predict the LP phase as the ground state observed experimentally, we report the numbers using the rev-vdW-DF2 functional of Hamada~\cite{Hamada14p121103} which is reported to capture the van der Waals interactions more accurately in other two-dimensional materials~\cite{Kim17p180101}. We also performed AIMD simulations using VASP for a bilayer 4 $\times$ 2 super-cell of CIPS, at 250K and 300K. We used different strengths for the out-of-plane electric-fields to investigate the switching dynamics. Trajectories as long as 35{\sl ps} were simulated to achieve good statistics and simulate rare switching events. 

\section{Results and discussion}

\begin{figure*}
\includegraphics[width=2.0\columnwidth]{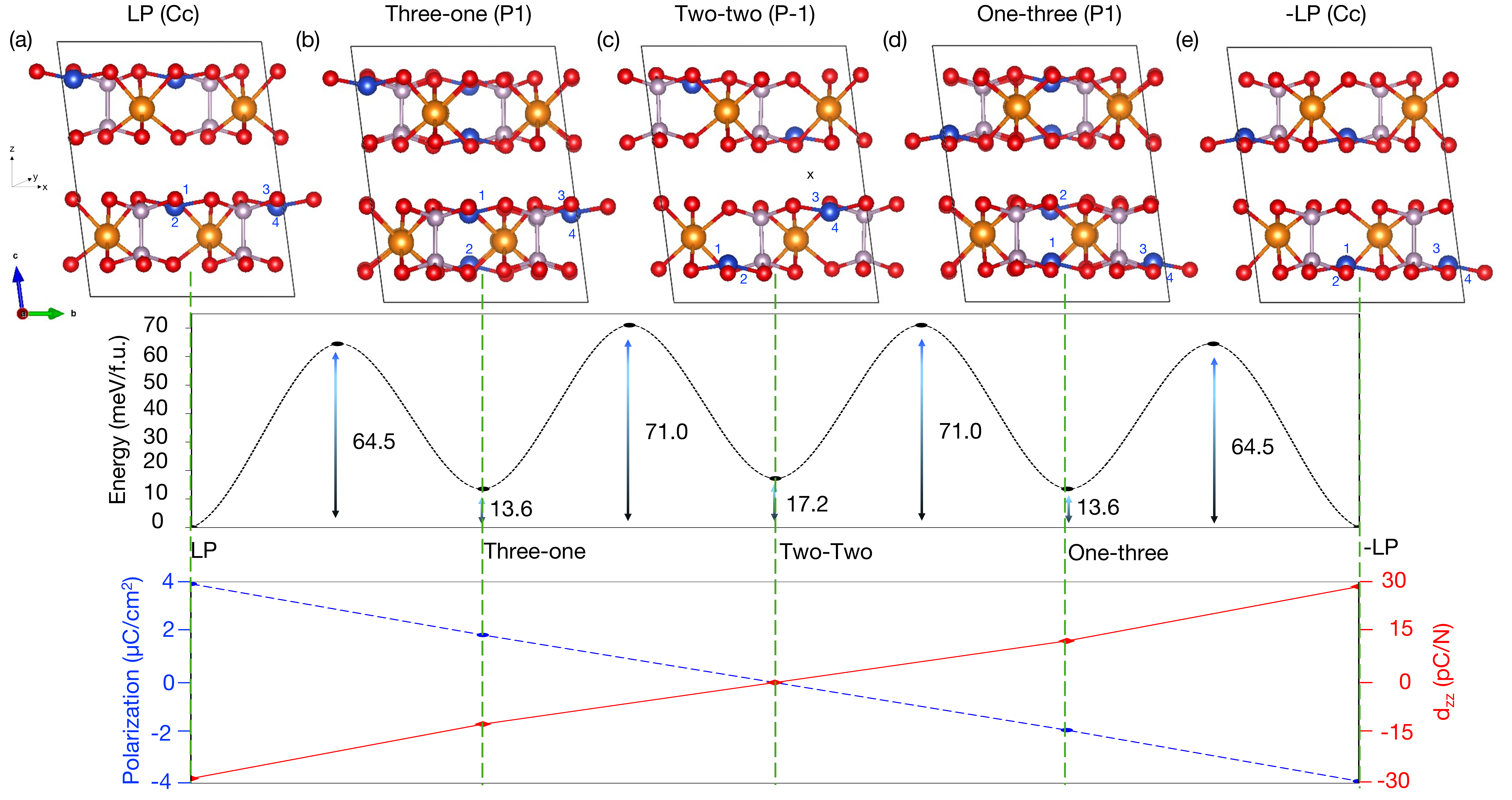}
\caption{\label{Fig2} The side view of the crystal structure of a) +LP phase, (b) the three-one structure, (c) the two-two structure, (d) the one-three structure, and (e) the -LP phase. The structures differ in the local position of the 4 Cu atoms (numbered blue balls) within each layer. The space group in each case is listed. The interlayer inversion center is marked by an `$\times$'. The stacking direction (z) differs from the crystallographic c-axis, and is also labeled. The total energy (black ellipses), the total polarization (blue ellipses) in ($\mu C/cm^2$) and piezoelectric-strain response $d_{33}$ (red diamonds) for the different structures are shown. }
\end{figure*}

 We investigate the ferroelectric switching of CIPS between the low-energy $\pm$ LP phases via the inversion symmetric reference paraelectric structure (PE) shown in Fig.~\ref{Fig1}. Uniform switching between up and down polarized ${Cc}$ phases can be achieved by a coherent motion of Cu-atoms between the -LP and the +LP phases via the ordered PE phase where all the Cu-atoms are in an octahedral center~\cite{Sivadas22p013094}. The PE phase corresponds to the maximum energy barrier phase for a coherent switching of out-of-plane dipoles. This PE phase is unstable with two zone-center instabilities corresponding to a polar instability ($\Gamma_2-$) leading to the LP phase (Fig.~\ref{Fig1} (b)) and an anti-polar instability ($\Gamma_2+$) leading to an interlayer antiferroelectric (AFE) phase (Fig.~\ref{Fig1} (d)). The energy gain for the AFE phase from the PE phase due to the anti-polar distortions is $\sim$ 235 meV/ f.u.. The AFE phase, which forms in space group P-1, is still higher in energy by 17.2 meV/f.u. than the polar LP ground-state that forms in space group $Cc$. This suggests that the interlayer polarization prefers ferroelectric order. 
 
 We also considered the HP phase discussed by Brehm et al.~\cite{Brehm20p43}. We found this to be a saddle point when plotting the total energy relative to the Cu displacement in the fully-relaxed stress-free case (see supplementary material for details). Given the large degree of inhomogenous strain observed in experiments across CIPS/IPS phases~\cite{Brehm20p43}, and the tunability of the energy-surface by such in-plane strain~\cite{Sivadas22p013094}, such a saddle point could be stabilized in the experiments. Later, we will discuss our AIMD results which shows a finite occupation of the HP sites for some Cu sites at elevated temperatures (250~K and 300~K). We find that CIPS demonstrates a strain-tunable quadruple well for each Cu-atom independently, and the material behaves as a sublattice melted system due to availability of more number of Cu-sites than stoichiometry would allow. This giving rise to high ionic motion even at 250~K in our AIMD simulations, explaining the experimentally observed conductivity well below T$_c$$\sim$315~K.  Nevertheless, the ground-state of CIPS is the LP phase with an energy barrier of 252.2 meV/f.u. for coherent switching of Cu-atoms within the layers.  

\subsection{Local polar order}

To understand the localization of dipoles and their couplings within the layer we considered a 2$\times$2 super-cell of the bulk unit cell containing 4 Cu atoms in each layer (see Fig.~\ref{Fig2}) with interlayer FE order. We considered structures with similar local out-of-plane polar distortions as in the LP phase but where the distortions have a phase difference along the in-plane directions (see supplementary for more details). Each of the configurations in Fig.~\ref{Fig2} from left to right differ by the switching of a single Cu-atom from the top to bottom within a layer. Studying the relative stability of these structures allows us to compare the coherent switching of polarization with an incoherent disordered switching of polarization under an externally applied electric field. 

Figure~\ref{Fig2} shows the space groups formed by the different phases, their total energy per f.u., the out-of-plane polarization and the corresponding $d_{zz}$ for the phases considered~\footnote{We report only the out-of-plane polarization as this is of our primary interest. LP phase also hosts an in-plane polarization which is allowed by symmetry.}. In the 2 $\times$ 2 super-cell, we find that the LP phase is the lowest energy structure with the three-one and the two-two structures higher in energy by only 13.6 meV/f.u. and 17.2 meV/f.u., respectively. We notice that the three-one, two-two and one-three phases form local minimum suggesting their metastability. The highest energy barrier to switch a single Cu-atom from up to down polarization site is at the octahedral center. While the actual switching path is not necessarily a straight line, as evidenced from our AIMD simulations, the energy barrier to switch a single Cu-atom can be obtained by considering the energy difference between any of the metastable structure and the corresponding state with a single Cu-atom in the octahedral center. Given that we have a 2 $\times$ 2 super-cell, and the metastable structure shown in fig.~\ref{Fig2} from left to right are from the LP to the –LP phase, with each intermediate phase differing by a single Cu-atom switching in each layer, the barrier for a single Cu-atom switching can be estimated to be $\sim$203.6-258 meV (i.e. 4 $\times$ the different heights ranging $\sim$ 50.9-64.5 meV/f.u. in Fig.~\ref{Fig2}.). The largest barrier is within 6meV/f.u. of the barrier to switch all the Cu-atoms simultaneously (252.2 meV/f.u.), suggesting that at even very low temperatures, field-induced switching in CIPS will proceed via incoherent Cu-atom switching. 

Interestingly, the polarization of the structures is linear with respect to the number of Cu-atoms that are switched.  The same is true for $d_{zz}$ as well. The LP phases has the largest value of polarization (3.92~$\mu C/cm^2$), in good agreement with experiments~\cite{Mat-Horiz}. The polarization values of the three-one structure is approximately half of that of the LP phase in amplitude (1.89~$\mu C/cm^2$). As the two-two structure has space group P-1 and is inversion symmetric, the total polarization vanishes. The one-three structure has a polarization value identical in amplitude to that of the three-one structure but opposite in direction. This clearly suggests that CIPS shows scale-free polarization, where the bulk polarization is simply an additive of the individual molecular dipole unit. The polarization corresponding to a single molecular f.u. is thus $\sim$ 1.0 $\mu C/cm^2$. 

\subsection{Total energy as a function of site occupation}
\begin{figure}
\includegraphics[width=1.0\columnwidth]{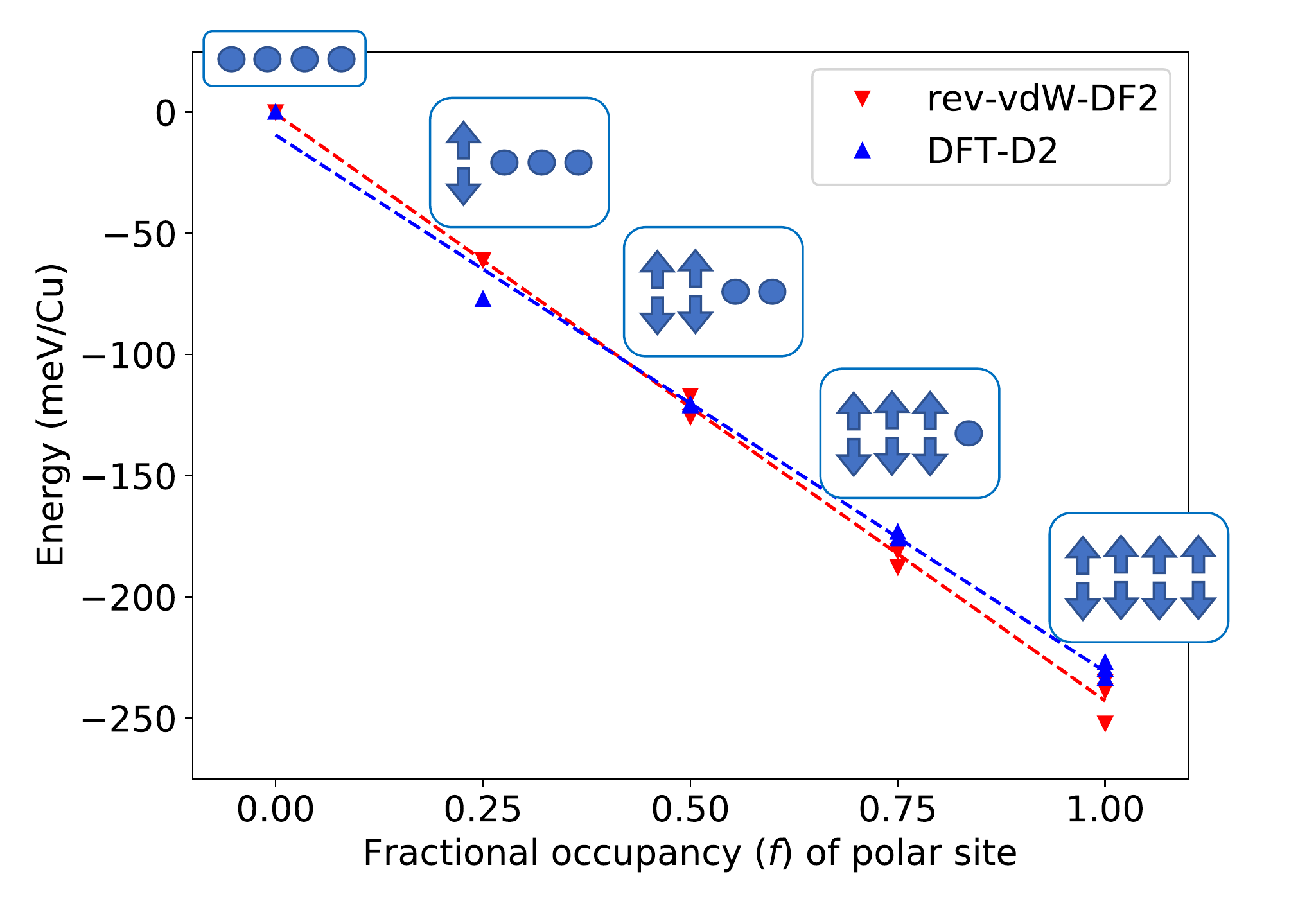}
\caption{\label{Efrac} The total energy (meV/f.u. atom) as a function of the fractional occupation ($f$) of the local polar sites. The schematic in the insert shows the local polar sites with up and down arrows within each layer. The filled circles represent the Cu atom in the local octahedral center. Many permutations of such sites with different directions for the local polar order are possible. The energy profile predominantly falls within a line.}
\end{figure}

To further check the scale-free nature of the polarization, we additionally consider a few cases where some fraction of the Cu atoms are in the local octahedral centers. The total energy for these different cases are plotted in Fig.~\ref{Efrac} as a function of the fractional occupation of the polar sites ($f$). The $f$ = 0 case corresponds to the PE phase (see Fig.~\ref{Fig1} (a)) with all the Cu atoms occupying the octahedral centers (shown in the insert as an array of filled circles). All the structures considered in Fig.~\ref{Fig2} (a)-(e) corresponds to $f$ = 1 where the Cu atoms occupy only the local polar sites. The insert shows a schematic of arrows pointing up and down to represent the many combinations of local polar ordering that is possible within each layer. $f$ = 0.75 corresponds to the case where only one of the Cu atom within each layer is in the octahedral center with the other three Cu atoms occupying the polar sites (represented by one filled circle and three up and down arrows). Incidentally, this also corresponds to the one-step switching barrier between the different $f$ = 1 structures in the 2$\times$2 super-cell. For instance, the switching from the LP in Fig.~\ref{Fig2} (a) to the three-one phase in Fig.~\ref{Fig2} (b) involves an intermediate state with one Cu atom per layer (labeled `2') is in the octahedral center half way between the Cu polar positions in Fig.~\ref{Fig2} (a) and Fig.~\ref{Fig2} (b). Similarly, the other $f$ = 0.75 cases corresponds to other one-step switching barrier between the different $f$ = 1 cases. 

The resultant energy profile in Fig.~\ref{Efrac} has a linear dependence on the fraction occupation of the polar sites. This shows that the total energy is largely additive and is primarily a function of the fractional occupation of the local polar sites. This is a direct result of our findings that $C$ in Equ.~\ref{Eq1} is much smaller than $V_{on-site}$. We thus prove that the individual dipoles are highly localized and as such in the thermodynamic limit should lead to a large number of disordered metastable states even at much lower temperatures than T$_c$, and even at finite electric fields. 

\subsection{Quantify the inter-dipole coupling and comparison with HfO$_2$}

To quantify the degree to which the dipoles are independent and localized we computed the onsite potential ($V_{on-site}$) as well as the inter-site coupling ($C$) as defined in Equ.~\ref{Eq1}. $V_{on-site}$ is the energy difference between the PE and LP phase. We previously discussed the importance of including the on-site anharmonic coupling of the polar mode and the fully symmetric mode to correctly characterize the energy surface~\cite{Sivadas22p013094}. Here, the $V_{on-site}$ has this local anharmonicity built into it. $C$ is the energy difference corresponding to an isolated flip of the oscillator. We define this as the energy difference between the LP phase and the three-one phase. We find $V_{on-site}$ and $C$ to be -252.2 meV/f.u. and 13.6 meV/f.u., respectively. The positive sign of $C$ signifies that the ordered LP phase corresponds to the lowest energy structure. The small $C$ to $V_{on-site}$ ratio justifies our prior conclusion that CIPS is a scale-free switchable ferroelectric with weakly coupled localized dipole molecular formula units. It also suggests that at finite temperatures where Cu-motion is activated, we should expect an incoherent switching mechanism of local dipoles on application of an electric field in CIPS rather than a coherent switching involving interlayer Cu hoping as previously proposed~\cite{Neumayer20p2001726}. Next, we discuss this incoherent switching path under an externally applied field. 

To investigate the scale-free switching of the ferroelectric under a finite electric field we revisit the total energy plot in Fig.~\ref{Fig2}. We compare the switching barriers between coherent switching and a local switching that is incoherent. The latter corresponds to the energy for switching a single Cu atom. The barrier to switch a single Cu-atom is 64.5 meV, and if there was no coupling between Cu-atoms, this will amount to 258 meV per f.u. (i.e. 64.5$\times$4 meV/f.u.).  This is similar in magnitude to the 252.2 meV/f.u. needed for coherent switching of all the Cu-atoms. So, coupling the Cu-atom motion only has a minimal gain in energy ($\sim$6 meV). Given that the energy to switch one Cu-atom independently, is similar to the energy for coherent switching we expect CIPS to show multi-state polarization with scale-free ferroelectric switching down to the nanometer scale, similar to the case of HfO$_2$~\cite{Lee20p1343}.

There are crucial differences between HfO$_2$ and CIPS. In CIPS, there is significant anharmonic contribution to the switching~\cite{Sivadas22p013094}, and the switchable polarization unit can be as small as a single f.u. (with a polarization of $\sim$ 1 $\mu C/cm^2$) without the need for a spacer non-polar layer. In fact, the barrier for the flipping of a single Cu atom is nearly independent of which Cu-site flips, suggesting that the size of any domain-wall in CIPS will also be atomically sharp. The absence of a spacer layer in CIPS effectively reduces the single-flip switching barrier to utmost ($\sim$ 258 meV), which is at least five times smaller than the single-flip barrier in Hafnia (1380 meV). This lower energy barrier should lead to lower power requirement for non-volatile switching.



\subsection{Mechanism of switching}

\begin{figure}
\includegraphics[width=1.0\columnwidth]{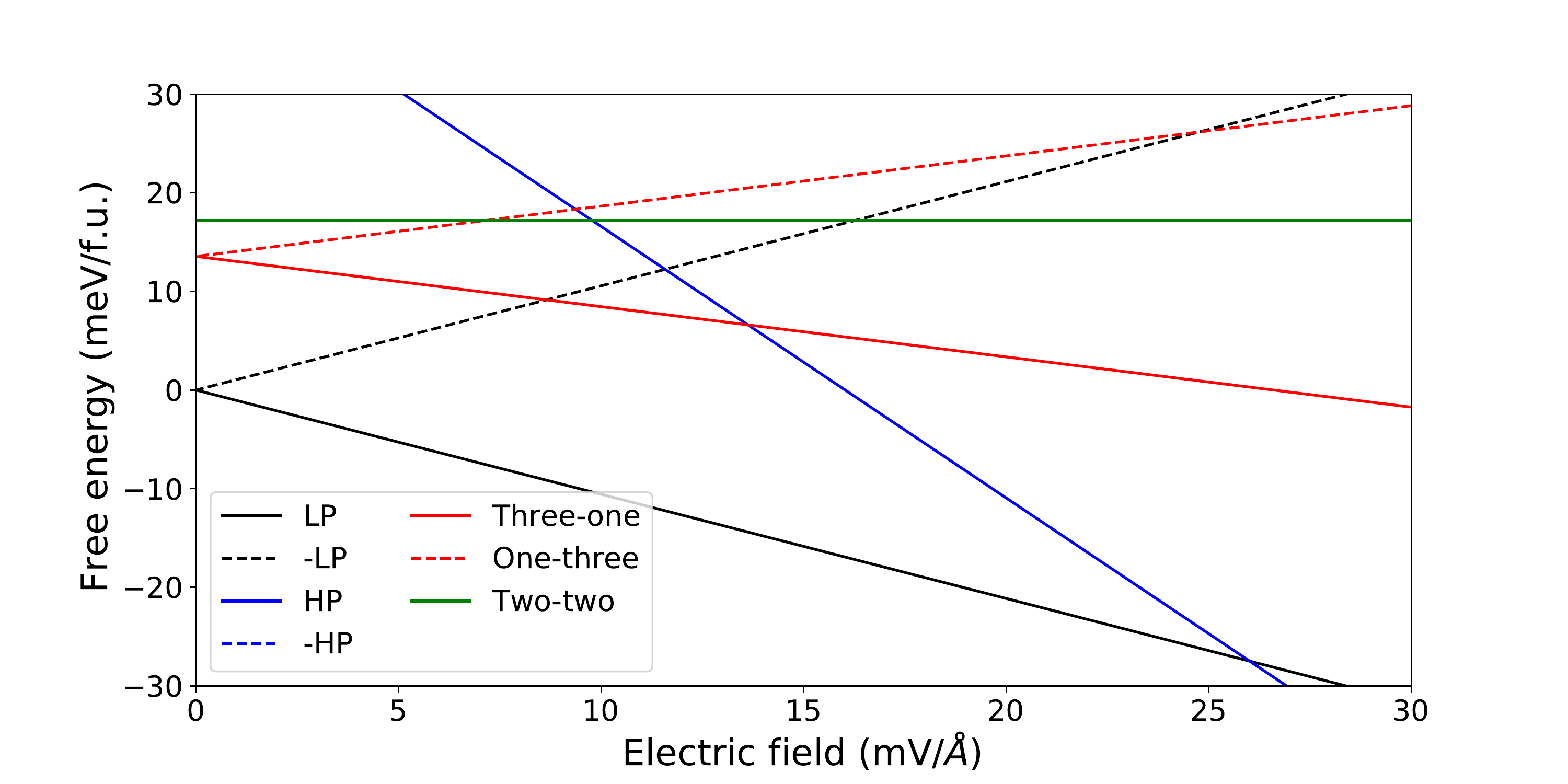}
\caption{\label{Fig-PD} The phase diagram under an electric field. Free energies for the different structures shown in Fig.~\ref{Fig2} and the HP phase. When the field is over 25 mV/\AA~ (when the dashed black and dashed red lines crosses), the -LP phase transitions over to one-three phase. One-three phase under this field is unstable and transition to the LP phase via the two-two phase and the three-one phase.}
\end{figure}

To understand the field-induced switching pathway, we look at the phase stability of the ground- and metastable-phases (as shown in Fig.~\ref{Fig2}) under an electric field \textbf{E}. We define the free energy of a system with an internal (zero-field) energy of V (for instance, in Equ.~\ref{Eq1}) to be F = V - \textbf{P} $\cdot$ \textbf{E}, where \textbf{P} is the polarization. For zero electric field both LP and -LP phases are isoenergetic, and form the degenerate ground states. Let us consider the scenario where initially the system is in the -LP phase. For a finite positive field the -LP phase becomes unstable, whereas the +LP phase becomes stable.  But due to a significant barrier of $E_b\sim$ 252.2 meV, the system remains in the metastable -LP phase. To fully overcome this barrier at room-temperature ($k_B T$ = 25.9 meV), one will require \textbf{E} = 107 meV/\AA $\sim$($(E_b - k_BT)/2P_{LP}$). But as we see from the phase-diagram in Fig.~\ref{Fig-PD}, as soon as we cross 25 mV/\AA~the one-three phase becomes relatively more stable than -LP. Given that the barrier for an individual Cu-atom to switch is similar to switching all the Cu-atoms, coupled with the fact that the two-two phase, the three-one phase and the LP phases are incrementally more stable than the one-three phase at E = 25 mV/\AA -LP phase should switch to the LP phase mediated by a weakly coupled switching of local dipoles.



\begin{figure}
\includegraphics[width=0.9\columnwidth]{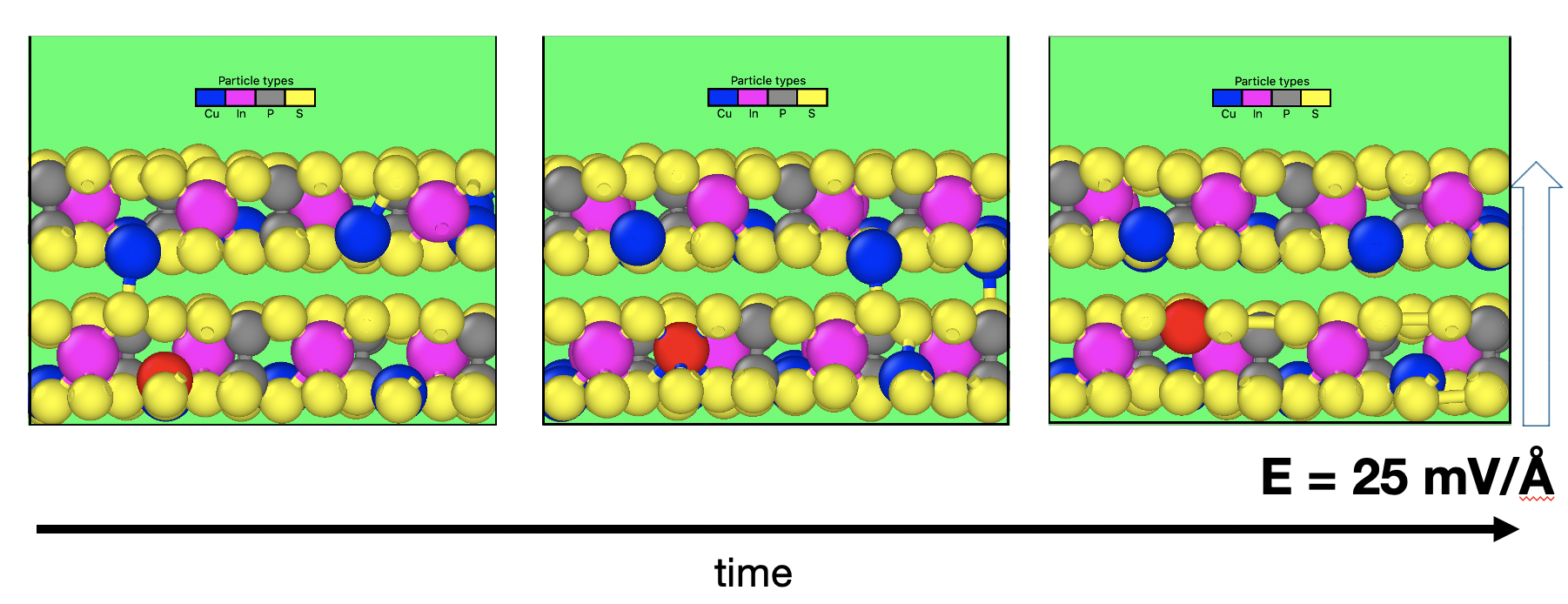}
\caption{\label{Fig-AIMD} Snapshots of atomic configurations in a portion of a 30 ps long {\sl ab initio} molecular dynamics (AIMD) trajectory of a CIPS bi-layer simulated with an out-of-plane electric field of 25 meV/\AA at T=250K. The `red' shaded Cu-atom is seen to move from the -LP site to the +LP site emphasizing the fact that CIPS switches by incoherent local dipole switching under finite fields.}
\end{figure}

We verify this by running AIMD simulations on the -LP phase of CIPS under an external electric filed near the critical value E = 25 mV/\AA (movie upload in SI). Fig.~\ref{Fig-AIMD} (a)-(c) shows atomic snapshots in a portion of the trajectory from a 30 ps long AIMD at 250~K, respectively. At $\sim$ 15 ps one Cu atoms (highlighted in red) in the bottom sub-layer of the bottom layer shifts to the top sub-layer of the bottom layer signifying a local change in polarization. This phase corresponds to the metastable one-three phase we previously discussed, consistent with the incoherent switching of local dipoles we propose. Upon increasing the temperature in the AIMD simulation to 300~K we find evidence of two Cu atoms switching events (see movie uploaded in SI). While longer AIMD simulations can show further evidence of complete switching mediated by the other phases we discussed, this is beyond the scope of our current computational campaign. Regardless, the AIMD simulations show clear signature of local incoherent switching of dipoles at E = 25 mV/\AA in CIPS, significantly below the critical field of 107 mV/\AA for coherent switching.

\begin{figure}
\includegraphics[width=1.0\columnwidth]{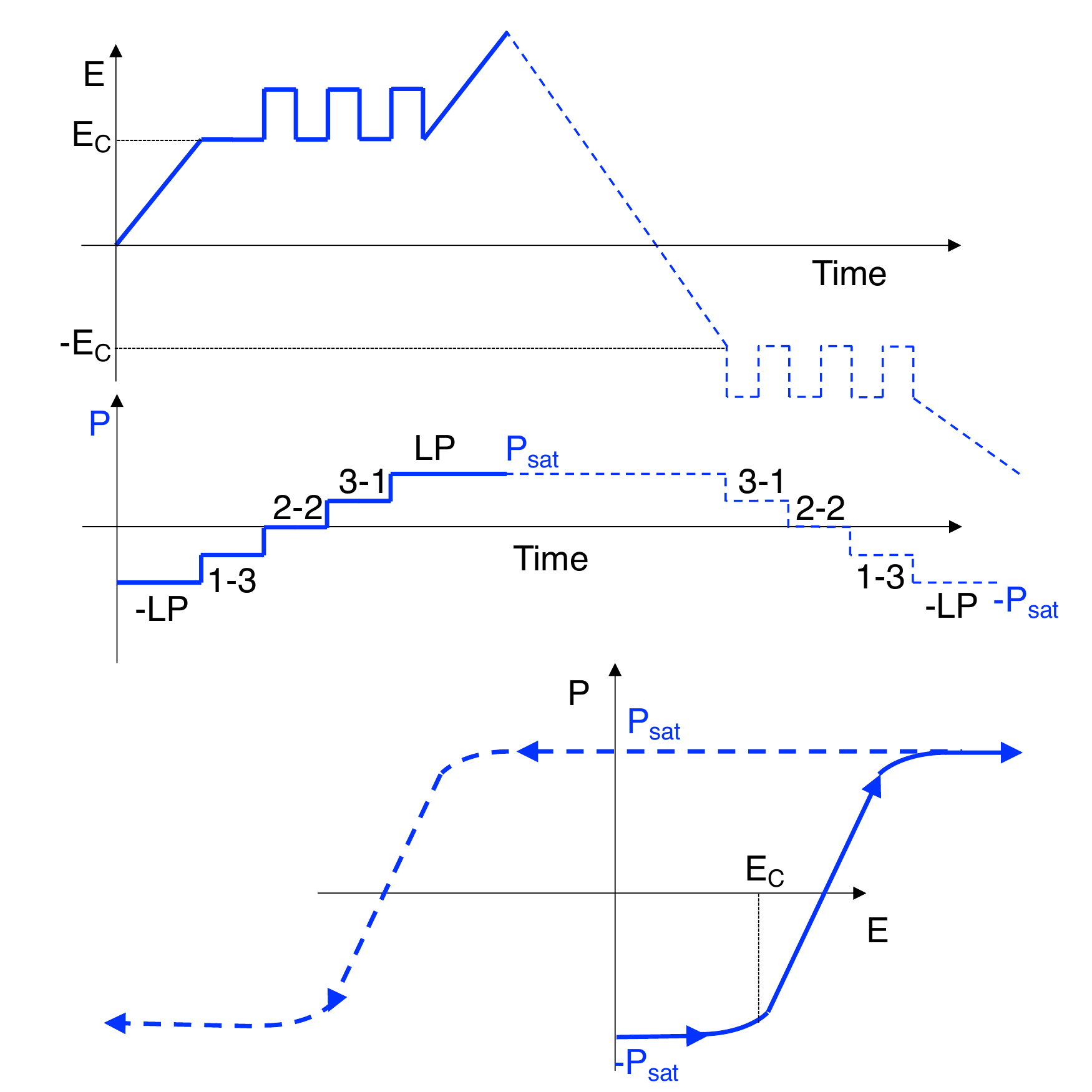}
\caption{\label{Fig-Ploop} Schematic of switching from the -LP phase to the LP phase under an applied field. (a) The Electric field pulse, (b) the polarization response to the field, and (c) the P-E loop are shown. Above the critical field, the -LP phase switches to the LP phase via many metastable phases.}
\end{figure}

Given the probability of switching one Cu-atom is the same as switching the entire bulk, the system effectively behaves as a non-interacting Ising-gas. As structures formed with different polar ordering have a barrier between them, these states will remain metastable when the field is below the critical field. As such, this continuum of metastable states can lead to memristive behavior, controlled by the local switching in response to a field. For example, a pulsed field of finite duration as shown by the schematic in Fig.~\ref{Fig-Ploop} can initially lead to a single Cu-atom switching from the -LP phase to a 1-3 phase in our super-cell model when the critical field is $E_c \sim$ 25 mV/\AA. This 1-3 metastable phase will remain stable once the field is removed due to the large barrier for switching a single Cu-atom. As shown schematically in Fig.~\ref{Fig-Ploop} consecutive pulsed fields on the structures can therefore lead to a cascade of single Cu-atom switching till we reach the +LP phase. In the thermodynamic limit, this translates to a cascade of continuum of metastable states, leading to a sloping hysteresis loop as we switch from the -LP to the +LP phase. This is indeed what was observed in experiments~\cite{Brehm20p43,Neumayer20p2001726, Mat-Horiz}, where switching between any two (strain-stabilized) long-lived phases had a sloping hysteresis loop. Availability of multitudes of kinetic pathways for switching as in a dipolar-glass will also give rise to a wake-up effect. The fact that switching in different parts of the same material do not couple with each other at the smallest unit-cell length, one could also envision a nanocapacitor with gates that independently tune different polar units in the same material, giving rise to ideal memristive behavior with large number of metastable polar states (see inset schematic in Fig. 4).  

The reported barrier for switching of the Cu atoms across the vdW gap via the HP phase is 850 meV when it is facilitated by vacancy formation~\cite{Susner17p7060}, which is three time larger than the in-plane switching barrier. 
Nevertheless, participation of the HP site in forming metastable structures at higher temperatures, as evidenced in prior experimental studies as well as our long AIMD simulations (not shown), should only favor incoherent switching within vdW layers due to more available sites for Cu-occupancy, and likely also across the vdW layers when sufficient Cu-vacancies are present. Overall, our results suggest a relatively weak coupling of local dipoles in CIPS compared to conventional ferroelectrics along with a relatively low barrier for switching of polarization compared to HfO$_2$, making it an ideal candidate to explore scale-free switching of polarization.

\section{Conclusion and Outlook}


We study the on-site anharmonicity as well as the intralayer inter-dipole coupling in CIPS. We find that the polar phase is stabilized by the anharmonic coupling between the polar mode and the Raman mode. This stabilization results in a weakening of inter-dipole coupling within each layer. The scale-free polarization (and related properties, such as piezoelectricity) and its incoherent field-induced switching even at low temperatures is a direct consequence of this weakening of inter-dipole coupling. 

Our investigation here is limited to the pristine crystalline phases of CIPS with no heterogeneity. In the presence of Cu vacancies it is reasonable to expect more in-plane switching paths for Cu atoms depending on the local environment. This could lead to additional incoherent switching pathways not considered here. Also, while our discussion is primarily focused on the change in the local occupation of Cu atoms within the layer, our MD simulations also show that the HP sites have a non-zero occupation at finite temperatures, and thus could additionally increase the degree of metastability in CIPS, thereby favoring incoherent field-induced switching at further lower temperatures.  

In summary, we find that CIPS behaves as a sublattice melted system due to availability of more number of Cu-sites than stoichiometry would allow, and tuning the energy of these sites by strain so they get closer to each other (e.g. to form a quadrupule well), or by incorporating additional low-energy Cu-sites by making CIPS Cu-deficient, will further favor incoherent field induced switching. This individual Cu-motion activation switching should naturally lead to large ionic conductivity as measured experimentally~\cite{Mat-Horiz}. Compared to HfO$_2$, one of the leading material candidates for scale-free polarization applications, CIPS requires at least five times smaller energy for the switching of polarization, allowing a scale-free switching at a smaller electric field than HfO$_2$. Further, due to the weak inter-dipolar coupling, we expect the domain walls to be atomically sharp, and not requiring a spacer layer such as in HfO$_2$. CIPS hence falls into a unique class of 2D ferro-ionic material with large ionic motion of Cu atoms that determines the ground state polar phases, ferroelectric switching and related field-induced responses. These findings are expected to spur new experiments to verify our predicted scale-free polarization and incoherent field-induced switching of CIPS, and incorporate CIPS in novel neuromorphic device geometries to realize low-power microelectronic applications.

\bibliography{citations-new}

\section{Supplementary information}

\subsection{Crystal structure of the LP phases}
\begin{table}
\caption{\label{tab1} The computed structural parameters for the LP phase of CuInP$_2$S$_6$ with space group $Cc$ are compared to the corresponding experimental values~\cite{Maisonneuve97p10860}. $a$, $b$ and $c$ are the lattice constants. The Wyckoff positions of the symmetry inequivalent atoms are also shown. The experimentally reported Wyckoff values are shown in brackets.}
\begin{tabular}{|c|c|c|c|} \hline
Lattice constants & DFT-D2 & Exp.\\
\hline
$a$ (\AA) & 6.10  & 6.10\\
$b$ (\AA) & 10.55  & 10.56\\
$c$ (\AA) & 13.80 &13.62 \\
$\beta$ ($^\circ$) & 107.32 & 107.10 \\
\hline
\end{tabular}
\begin{tabular}{|c|c|c|c|c|c|} \hline
Atom & Wyckoff  & $x$ & $y$ & $z$ \\
 & site & & & \\
\hline
Cu1 & 4a &  0.583 (0.596) & 0.336 (0.336) & 0.368 (0.389)\\
In & 4a &   0.500 (0.500) & 0.002 (0.002) & 0.252 (0.252) \\
P1 &  4a &   0.069 (0.069) & 0.168 (0.169) & 0.350 (0.351)\\
P2 &  4a &   0.951 (0.951) & 0.167 (0.167) & 0.180 (0.180)\\
S1 &  4a &   0.780 (0.781) & 0.155 (0.151) & 0.397 (0.397)\\
S2 &  4a &   0.741 (0.733) & 0.164 (0.165) & 0.895 (0.895)\\
S3 &  4a &   0.283 (0.285) & 0.015 (0.018) & 0.398 (0.397)\\
S4 &  4a &   0.243 (0.240) & 0.177 (0.173) & 0.138 (0.135)\\
S5 &  4a &   0.248 (0.256) & 0.178 (0.175) & 0.642 (0.642)\\
S6 &  4a &   0.281 (0.272) & 0.498 (0.494) & 0.641 (0.640)\\
\hline
\end{tabular}
\end{table}

While Maisonneuve et al.~\cite{Maisonneuve97p10860} and Qi et al.~\cite{Qi21p217601} reported an out-of-plane lattice constant of 13.62~\AA and 13.76~\AA, respectively, others have reported a smaller out-of-plane lattice constant of 13.19~\AA~\cite{Brehm20p43}. This discrepancy arises from the lack of unique choice of unit cell. 


\subsection{High-polarization (HP) phase}

\begin{figure}
\includegraphics[width=1.0\columnwidth]{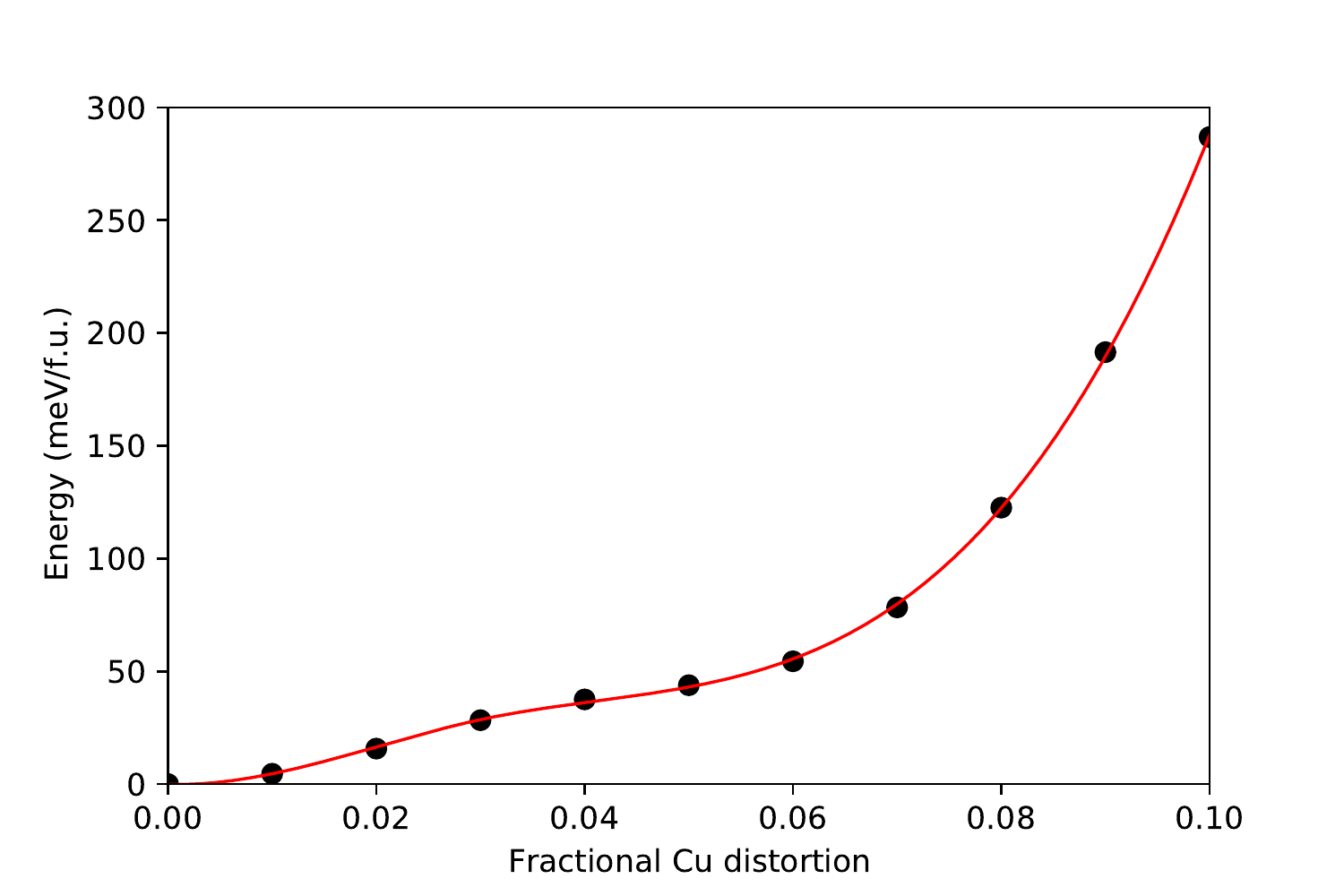}
\caption{\label{HP-phase} The total energy per formula units as the fractional position of the Cu atom is changed relative to the low-polarization (LP) phase.}
\end{figure}

Fig.~\ref{HP-phase} shows the total energy per formula units as the fractional position of the Cu atom is changed relative to the low-polarization (LP) phase. The energies are reported after full structural relaxation constraining the fractional position of the Cu atoms relative to the center of mass of the P atoms. The HP phase corresponds to the saddle point in the energy curve at a fractional distortion of 0.5 in

\subsection{Structures in the 2 $\times$ 2 super-cell}

As there are 8 Cu atoms in the 2 $\times$ 2 super-cell there are many structural combinations to consider. We initially limit ourselves to structures with only two unique local Cu positions, with either the Cu atoms near the top S layer or near the bottom S layer. Later, when we compute the total energies as a function of the fractional occupation we additionally consider a third local site where the Cu atoms can occupy, i.e., in the octahedral center within the layer similar to the PE phase. 

Fig.~\ref{Fig2} (a)-(e) shows the five structures considered where the Cu atoms are at different local polar sites within the 2 $\times$ 2 super-cell. Fig.~\ref{Fig2} (a) and (e) corresponds to the +LP case and the -LP case respectively, with the polarization pointing up and down, respectively. Fig.~\ref{Fig2} (b) corresponding to the case where three Cu atoms within a layer are near the top S sub-layer within each layer and the other Cu atom (labeled `2') is near the bottom S sub-layer. This still leads to four interlayer Cu orderings. However, we ignore the interlayer ordering in our discussion as we expect the effect of it to be small. Fig.~\ref{Fig2} (b) shows one such combination. We refer to this phase as the three-one structure. Fig.~\ref{Fig2} (c) shows the case where two Cu atoms (labeled `3' and `4') are near the top S sub-layer within each layer and the other two Cu atoms (labeled `1' and `2') are near the bottom S sub-layer. We refer to this structure as the two-two structure. Again, the interlayer ordering is ignored. Fig.~\ref{Fig2} (d) corresponding to the case where one Cu atom (labeled `2') is near the top S sub-layer within each layer and the other three Cu atoms are near the bottom S sub-layer. We refer to this phase as the one-three structure. The one-three structure is the inversion partner of the three-one structure relative to the PE phase. So it has identical energy to the one-three phase but with the polarization direction reversed. Within this notation, the LP (-LP) phase corresponds to the four-zero (zero-four) structure. We can see that the series of figures from Fig.~\ref{Fig2} (a)-(e) shows one possible sequence of switching path from the +LP phase in (a) to the -LP phase in (e), with all the Cu atoms staying within the layer.

\subsection{Framework for computing piezoelectric response}
Prior experimental works on CIPS report the piezoelectric strain tensor components~\cite{Liu16p12357,Brehm20p43}. The piezoelectric stress tensor ($e$) and the piezoelectric strain tensor ($d$) are defined as
\begin{equation}\label{e1}
\begin{split}
    e_{i j} = \frac{\partial P_i}{\partial \eta_j} \bigg\vert_E, \\
    d_{i j} = \frac{\partial P_i}{\partial \sigma_j} \bigg\vert_E,
\end{split}
\end{equation}
drespectively, where $P_i$ is the Cartesian component of polarization that changes on application of a strain ($\eta$) and stress ($\sigma$).  We report the piezoelectric tensors in the Cartesian coordinates. $\eta_{z}$ is defined as ($c_z/c^0_z - 1$), where $c_z$ is the out-of-plane component of the c-lattice vector. We computed the piezoelectric stress tensor $e_{zz}$ by computing the change in polarization with respect to out-of-plane strain $\eta_{z}$, keeping the in-plane lattice constants fixed to the zero-stress case. To compare directly with experiments we additionally computed the piezoelectric stress tensor $d_{zz}$ using the value of stress after relaxing the in-plane lattice constants~\cite{Kim19p104115}. Another approach would be to compute $d_{zz}$ using
\begin{equation}\label{stress-strain}
d_{zz} = e_{zz}/C_{zz},
\end{equation}
where $C_{zz}$ is the Young's modulus. $C_{zz}$ is defined as
\begin{equation}\label{YM}
    C_{zz} = \frac{1}{V} \frac{\partial^2 E}{\partial^2 \eta_{z}}, 
\end{equation}
where $E$ is the total energy including the effect of internal strain, and $V$ the volume. We verified that these two methods yield similar results, and report the former values. Evidently from Eq.~\ref{e1}, both $e$ and $d$ are odd-parity responses and require an inversion asymmetric structure to be non-zero.

\subsection{Contributions to the piezoelectric response tensor}

To compare the different contributions to the PRs we computed the polarization as a function of strain $\eta_{z}$ for three cases: (a) clamped-ion (CI) case where neither the atoms nor the cell is relaxed, (b) relaxed ion (RI) case where the atoms are relaxed while constraining the in-plane lattice constants, and (c) full relaxation (FR) where both the atoms and the in-plane lattice parameters are relaxed. Such a decomposition has been used to distinguish the clamped-ion contribution from the effect of the internal atomic relaxation in response to the macroscopic strain~\cite{Liu17p207601}. 


The CI contribution is sometime refereed to as the frozen-ion contribution as the fractional amplitude of the ions are fixed when the strain is applied. The RI component of the piezoelectric stress tensor ($e$) accounts for the effect of the internal strain as the polarization changes due to atomic distortions. In this case, the proper and improper responses are identical and unambiguously defined independent of the choice of the branch of polarization~\cite{Vanderbilt00p147,Dreyer16p021038}. We compared the computed values of $e_{zz}$ with the DFPT method implemented with VASP and found excellent agreement for the LP phase. 

As the piezoelectric strain tensor ($d$) is defined under zero-stress boundary condition, $d_{zz}$ for the FR case leads to the correct PR~\cite{Kim19p104115}. There is a distinction between the improper component of $d_{zz}$ and the proper component of $d_{zz}$~\cite{Kim19p104115}. The former describes the change in polarization with respect to stress, whereas the latter includes the effect of change in area as well. As a stress-induced current is measured in most experimental setups, a change in unit-cell area can have a significant contribution. Therefore the proper component of $d_{zz}$ is the appropriate value to compare with experiments. However,  as the effect of the relaxation of the in-plane lattice constants for the different values of out-of-plane strain $\eta_{zz}$ is negligible in this class of materials, we report only the proper value. This is also consistent with the previous reports of a close to zero Poisson's ratio in this family of materials~\cite{Qi21p217601}.

\subsection{Young's modulus}

\begin{figure}
\includegraphics[width=1.0\columnwidth]{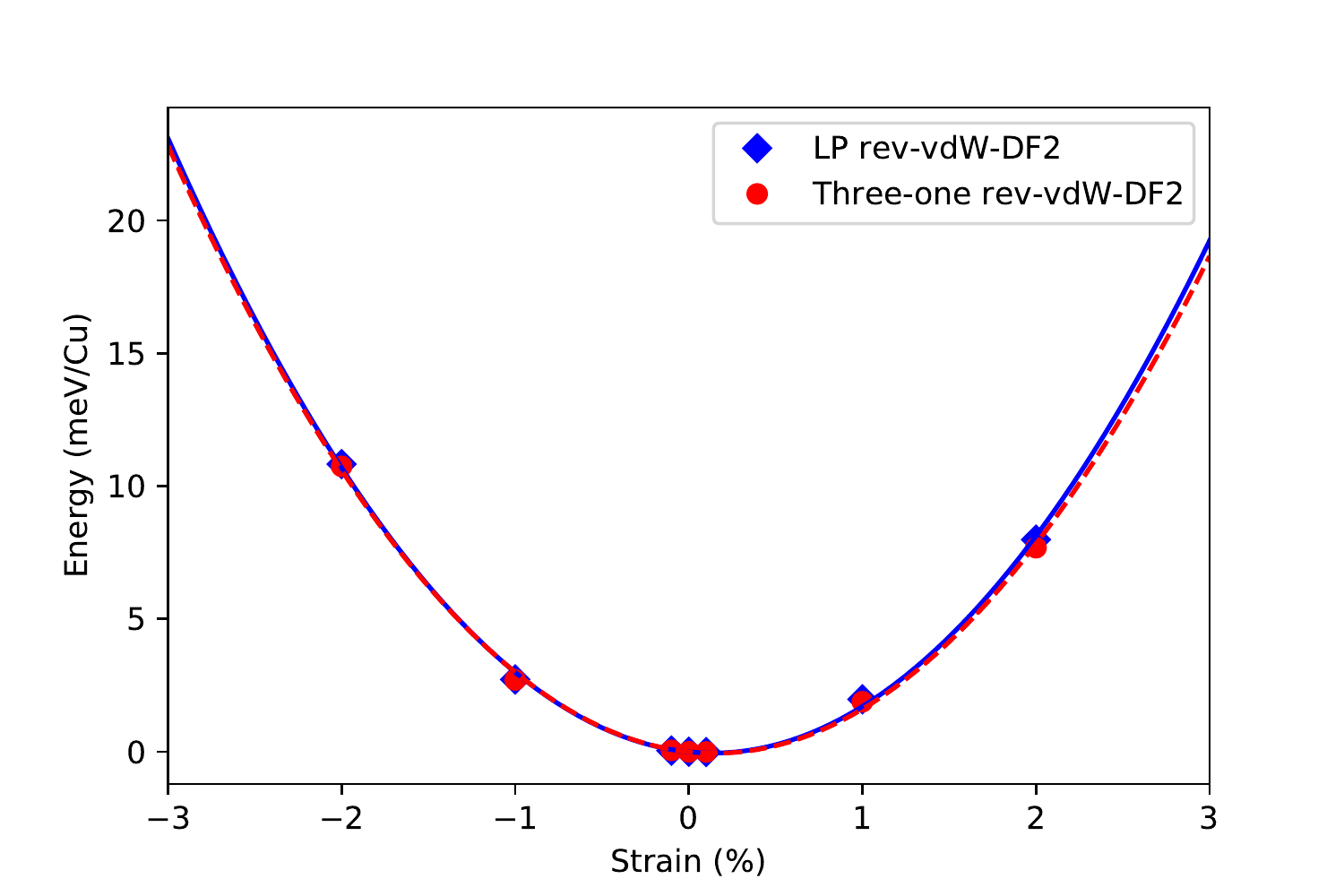}
\caption{\label{Young} The total energy per Cu atoms as a function of strain-percentage for the LP phase (blue diamonds) and the three-one structure (red dot) for rev-vdW-DF2 functional.}
\end{figure}

Fig.~\ref{Young} shows the total energy per Cu atoms as a function of strain-percentage for the LP phase and the three-one structure for rev-vdW-DF2. The results are similar with DFT-D2. The energy per Cu atom is indicative of the total energy per unit volume, and shows that the Young's modulus ($C_{zz}$) for the two structures are similar.

\subsection{Piezoelectric response}

\begin{figure*}
\includegraphics[width=2.0\columnwidth]{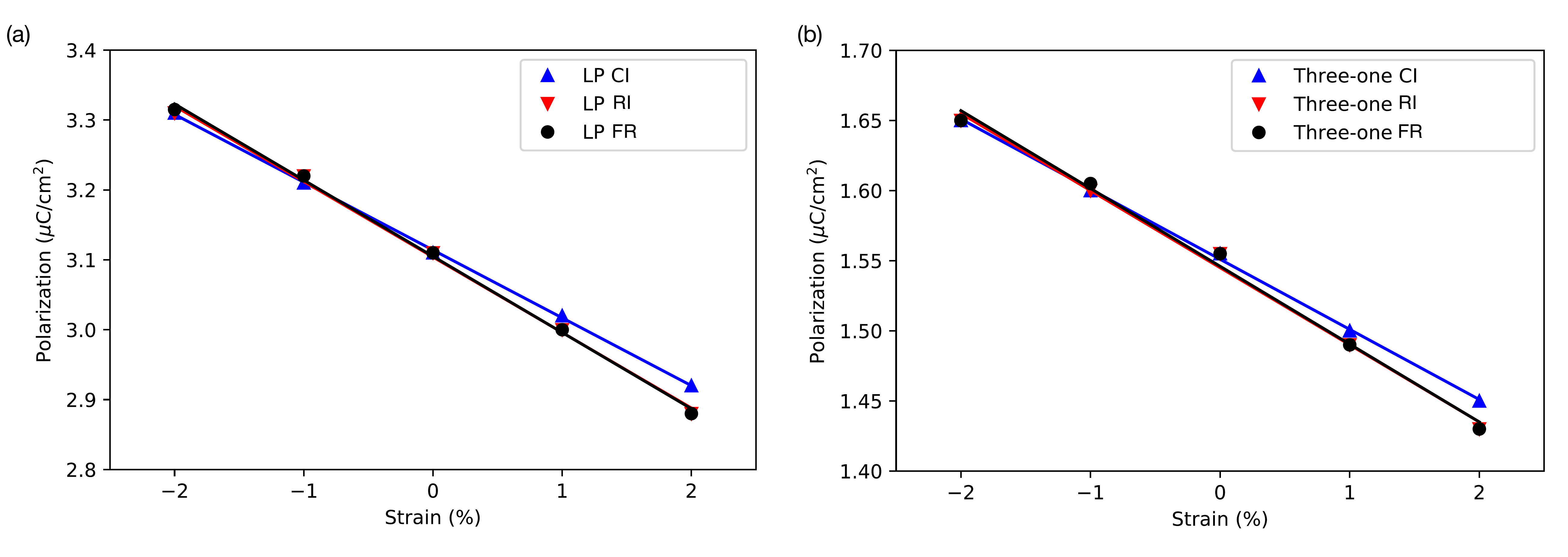}
\caption{\label{Fig3} The total polarization in ($\mu C/cm^2$) as a function of strain percentage for (a) the LP phase and (b) the three-one phase. The three cases for the clamped-ion (CI) case (blue upward triangle), the relaxed-ion (RI) case (red downward triangle), and fully-relaxed (FR) case (black dotted) are shown. The slope for the RI case yields the piezoelectric-stress response, $e_{zz}$.}
\end{figure*}

We also computed the PRs for the different structures considered by computing the change in polarization on application of strain and stress. Fig.~\ref{Fig3} shows the polarization for (a) the LP phase and (b) the three-one structure as a function of strain. This is shown for the CI case (blue triangles), for the RI case (red downward triangles), and for the FR case (black circles). The RI case yields the relevant piezoelectric-stress coefficient $e_{zz}$. The change in polarization with respect to strain for the two-two structure is zero, as expected from symmetry, and hence not shown. The polarization response for the one-three structure can be obtained from the three-one structure by reversing the sign of the polarization.

\begin{figure}
\includegraphics[width=1.0\columnwidth]{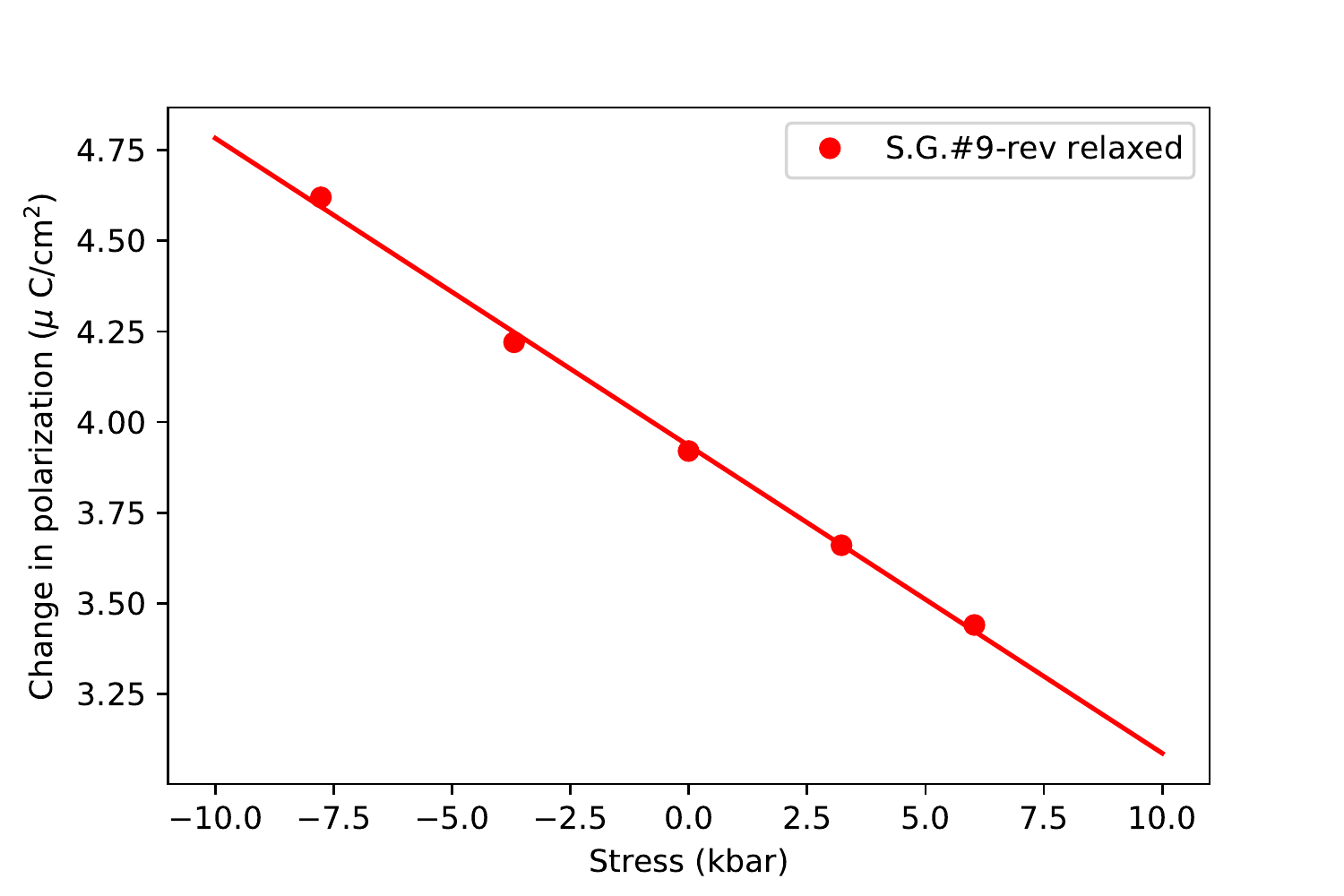}
\caption{\label{P-strain} The total polarization in ($\mu C/cm^2$) as a function of stress in kbar for the LP phase. The slope yields the piezoelectric-strain response, $d_{zz}$. }
\end{figure}

Fig.~\ref{P-strain} shows the total polarization as a function of stress for the LP phase. The slope yields the piezoelectric-strain response. 

\begin{figure}
\includegraphics[width=1.0\columnwidth]{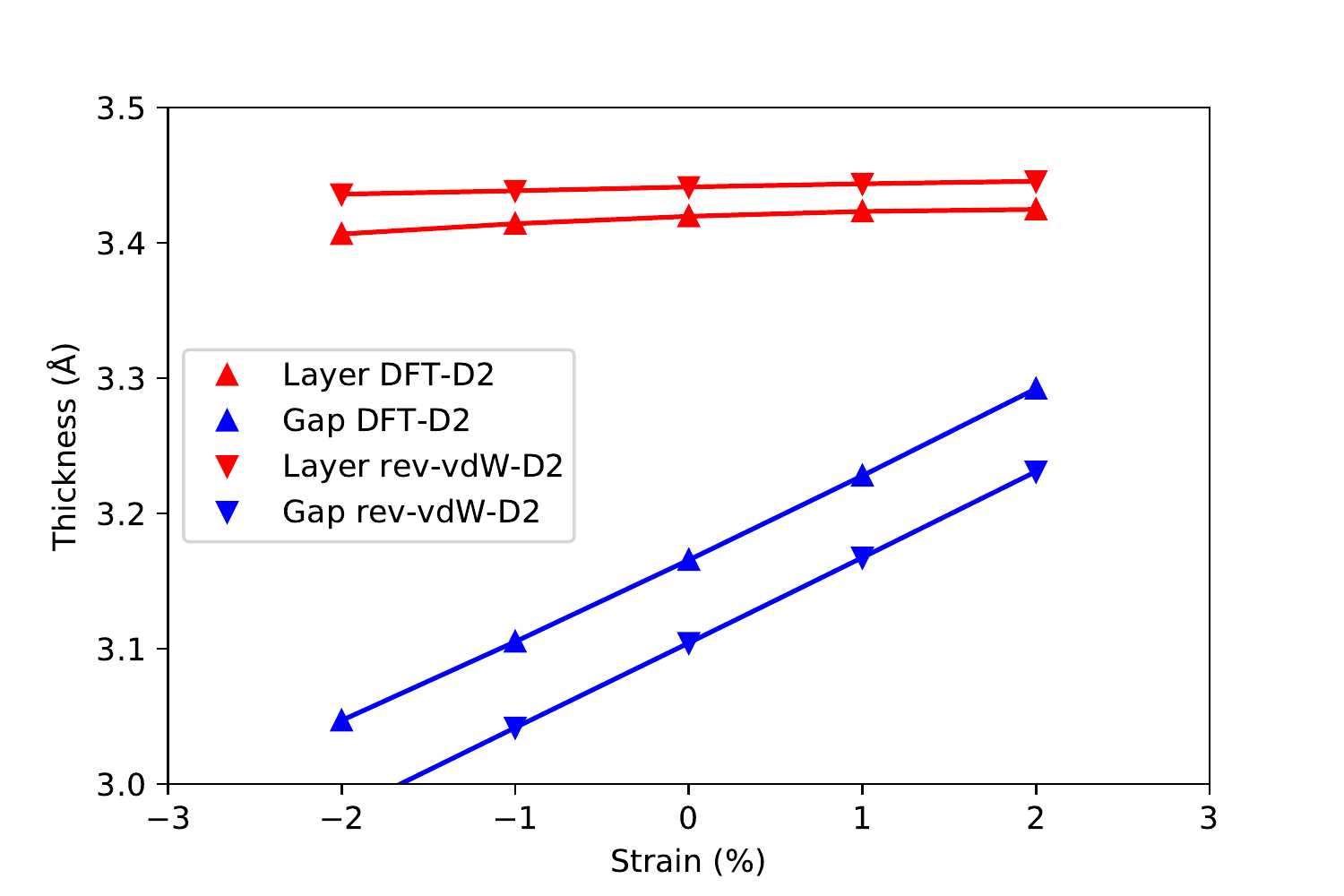}
\caption{\label{tLG} Changes of interlayer gap length (blue) and layer thickness (red) with respect to strain for the different functionals. While they show similar qualitative trends, rev-vdW-DF2 functional creates a thicker CIPS layer with less vacuum spacing between the layers. }
\end{figure}


When analyzing the individual components to $e_{zz}$ Fig.~\ref{Fig3} shows that there is no noticeable difference between the CI, the RI and the FI cases. We conclude that the change in polarization with strain is driven primarily by the negative clamped-ion (CI) term of the PR for both structures. The contribution of the internal strain is small, consistent with Qi et al.~\cite{Qi21p217601}. The effect of the internal strain is predominantly to changes the interlayer gap (see Fig.~\ref{tLG}). Finally, by comparing the FR case and the RI case we find that the effect of the out-of-plane strain on the in-plane lattice relaxation is minimal, consistent with the reportedly close to zero Poisson's ratio in CIPS~\cite{Qi21p217601}. 


The polarization as a function of strain for the LP phase compares well with prior results~\cite{Neumayer19p024401, Qi21p217601}. The $e_{zz}$ of -10.8 $\mu C/cm^2$ that we find with DFT-D2 for the LP phase also compares well with the reported -9.7 $\mu C/cm^2$ in Qi et al.~\cite{Qi21p217601} and -13.7 $\mu C/cm^2$ in You et al.~\cite{You19p3780} who used DFT-D2. However, rev-vdW-DF2 yields $e_{zz}$ of -28.6 $\mu C/cm^2$ for the LP phase demonstrating a strong dependence on the choice of functional for $e_{zz}$. 
Nevertheless, when we compare the polarization as a function of strain for the three-one structure we found that $e_{zz}$ is approximately half in amplitude as that of the LP phase for both DFT-D2 (-5.5 $\mu C/cm^2$) and rev-vdW-DF2 functionals (-12.4 $\mu C/cm^2$). This suggests that similar to polarization, $e_{zz}$ also depend primarily on the local polar off-centring of the Cu atoms, similar to the total polarization. They also scale with the total polarization. 

As prior experiments report the piezoelectric strain tensor ($d_{zz}$), we use Equ.~\ref{stress-strain} to convert $e_{zz}$ to $d_{zz}$. Therefore, we computed the Young's modulus $C_{zz}$ using Equ.~\ref{YM}. A quadratic fit to the energy yields $C_{zz}$ 35~GPa and 31~GPa for the LP phase and three-one structure, respectively, for DFT-D2 functional. For rev-vdW-DF2 functional, the corresponding values were 36~GPa and 34~GPa, respectively. It is worth noting that our computed $C_{zz}$ is significantly different from the first-principles results for the LP phase (7.5~GPa) reported in You et al.~\cite{You19p3780}. However, it compares well with the experimental reported values of 25 GPa listed in the same work~\cite{You19p3780}. The resultant $d_{zz}$ values for the different phases and the different functionals are summarized in Table~\ref{tab2}. A strong dependence of the choice of functional is evident. The $d_{zz}$ values for rev-vdW-DF2 functional (-8.0 pC/N) is closest to the experimental reported values of -11.8 $\pm$ 1.3 pC/N for the +LP phase~\cite{Brehm20p43}. 

\subsection{Choice of functional}

\begin{table*}

\caption{\label{tab2} The space group of the different structures considered, their total energy (meV/f.u.) relative to the LP phase, the total polarization along the stacking direction ($\mu C/cm^2$), the piezoelectric stress tensor component $e_{zz}$ ($\mu C/cm^2$), and the proper piezoelectric strain tensor component $d_{zz}$ (pC/N). The results for for rev-vdW-DF2 functional (DFT-D2 in brackets) are shown.}
\begin{tabular}{|c|c|c|c|c|c|} \hline
Structure & Space Group & Total Energy & $P_z$ & $e_{zz}$ & $d_{zz}$\\
\hline
PE & C2/c & 252.2 (226.9) & 0 & 0 &  0 \\
AFE & P-1 & 17.2 (10.7) & 0 & 0 &  0 \\
LP & Cc & 0 (0) & 3.92 (3.11) & -28.6 (-10.8) & -8.5 (-3.1) \\ 
three-one & P1 & 13.6 (-2.9) & 1.89 (1.56) & -12.4 (-5.5) & -3.7 (-1.7) \\ 
two-two & P-1 & 17.2 (-6.5) & 0 & 0 & 0 \\
\hline
\end{tabular}
\end{table*}

We also find that the ground state configuration is sensitive to the choice of the functional used. The results from DFT-D2 is listed in Table~\ref{tab2} in brackets. For DFT-D2 functional, we find that the two-two structure shown in Fig.~\ref{Fig2} (c) has the lowest energy. This is in contrast to the case of the Selenides where the LP phase is the lowest energy structure~\cite{Brehm20p43, Sivadas22p013094}. The three-one and the one-three structures have energy intermediate to that of the two-two structure and the LP phases. Although, the energy difference between the differed structures shown in Fig.~\ref{Fig2} are small (6.5 meV/f.u.), and comparable to the thermal energy. As the experimental ground state corresponds to the LP phase, we use the rev-vdW-DF2 results to discuss quantitative trends from now on. We also provide the results for DFT-D2 to compare with prior theoretical works~\cite{You19p3780,Qi21p217601}.

Irrespective of the lowest energy structure, our results highlight that there are multiple structures with similar energies that are metastable. A preconditioning as reported in the experiments where a large positive voltage is applied to begin with will likely order the local dipoles, thereby stabilize the LP phase as the initial structure~\cite{Neumayer20p064063}. 

\end{document}